# Revealing the orbital origins of exotic electronic states with Ti substitution in kagome superconductor CsV$_3$Sb$_5$

Zihao Huang[1,2,*], Hui Chen[1,2,*,§], Hengxin Tan[3,*], Xianghe Han[1,2], Yuhan Ye[1,2], Bin Hu[1,2], Zhen Zhao[1,2], Chengmin Shen[1,2], Haitao Yang[1,2], Binghai Yan[3,†], Ziqiang Wang[4], Feng Liu[5], Hong-Jun Gao[1,2,‡]

[1] *Beijing National Center for Condensed Matter Physics and Institute of Physics, Chinese Academy of Sciences, Beijing 100190, China*
[2] *School of Physical Sciences, University of Chinese Academy of Sciences, Beijing 100190, China*
[3] *Department of Condensed Matter Physics, Weizmann Institute of Science, Rehovot, Israel*
[4] *Department of Physics, Boston College, Chestnut Hill, MA 02467, USA*
[5] *Department of Materials Science and Engineering, University of Utah, Salt Lake City, Utah 84112, USA*

**ABSTRACT**. The multiband kagome superconductor CsV$_3$Sb$_5$ exhibits complex orbital textures on the Fermi surface, making the orbital origins of its cascade of correlated electronic states and superconductivity a major scientific puzzle. Chemical doping of the kagome plane can simultaneously tune the exotic states and the Fermi-surface orbital texture, and thus offers a unique opportunity to correlate the given states with specific orbitals. In this Letter, by substituting V atoms with Ti in kagome superconductor CsV$_3$Sb$_5$, we reveal the orbital origin of a cascade of its correlated electronic states through the orbital-resolved quasiparticle interference (QPI). We analyze the QPI changes associated with different orbitals, aided by first-principles calculations. We have observed that the in-plane and out-of-plane vanadium $3d$ orbitals cooperate to form unidirectional coherent states in pristine CsV$_3$Sb$_5$, whereas the out-of-plane component disappears with doping-induced suppression of charge density wave and global electronic nematicity. In addition, the Sb $p_z$ orbital plays an important role in both the pseudo-gap and superconducting states in CsV$_3$Sb$_5$. Our findings offer new insights into multiorbital physics in quantum materials which are generally manifested with intriguing correlations between atomic orbitals and symmetry-encoded correlated electronic states.

[*] These authors contributed equally to this work.
Correspondence author: hchenn04@iphy.ac.cn, binghai.yan@weizmann.ac.il, hjgao@iphy.ac.cn

Orbital is a key degree of freedom in quantum materials, where multiple atomic orbitals contribute to their symmetry-breaking low-energy physics and unique properties [1–7]. Strongly-correlated materials with complex orbital textures at the Fermi surface (FS) exhibit intriguing orbital-dependent phenomena, such as orbital-dependent band renormalization [8,9], symmetry-breaking states [10,11], and orbital-selective Mott transition [12] and Cooper pairing [13]. Understanding the orbital nature of electronic states is thus essential for uncovering the mechanisms behind these phenomena.

The newly discovered kagome superconductor $A$V$_3$Sb$_5$ ($A$=K, Rb, Cs) is a multiband superconductor with complex orbital textures at the Fermi level, exhibiting a cascade of correlated electronic states, such as Z$_2$ topology [14–16], rotation-symmetry-breaking charge density waves (CDW) [17–20], time-reversal-symmetry-breaking states [21–24], pair density waves (PDW) [25] and electronic nematicity [26–28]. Also, $A$V$_3$Sb$_5$ has been found to exhibit rich orbital-dependent physics [29–32]. Specifically, the anisotropy of Knight shift in nuclear magnetic resonance measurements of CsV$_3$Sb$_5$ indicates possible orbital ordering and fluctuations [33]. Orbital-dependent carrier-doping effect is observed by doping electrons into the surface of CsV$_3$Sb$_5$ [34]. The CDW gap opens at FS of V $d$ orbital while leaving the Sb orbital intact [35,36], likely due to FS instabilities of multiple van Hove singularities from distinct V orbital-derived bands [37,38]. Additionally, it exhibits possibly multiband superconductivity [39,40], with the Sb $p_z$ orbital making seemly important contributions [41,42]. However, our fundamental understanding of multiple orbital physics in the $A$V$_3$Sb$_5$ kagome system is still far from complete, especially the orbital origins of different exotic electronic states remain largely unknown. This is mainly caused by a complex multiorbital FS [32,35,36,38] and the existence of multiple intertwined states [18,26,28,43], which make it difficult to correlate one-by-one the former with the latter.

Recently, chemically doped $A$V$_3$Sb$_5$ kagome systems have been demonstrated to effectively tune the multiple electronic states [40,42,44,45]. Meantime, the chemical doping is expected to also change the orbital textures at the FS. Therefore, we are motivated to investigate the evolution of Fermi-surface orbital textures along with disentangling the cascade of correlated states at different doping levels, so that the correlation between the two can be resolved. In this Letter, we systematically substitute V with Ti in kagome superconductor $CsV_3Sb_5$, and perform the orbital-resolved quasiparticle interference (QPI) measurements at low temperatures. We have revealed the orbital origins of all the observed quantum states, state-by-state and orbital-by-orbital, by analyzing the QPI spectra in combination with scanning tunneling microscopy/spectroscopy (STM/STS), aided with density-functional theory (DFT) calculations (details in Supplemental Material Ref. [46]).

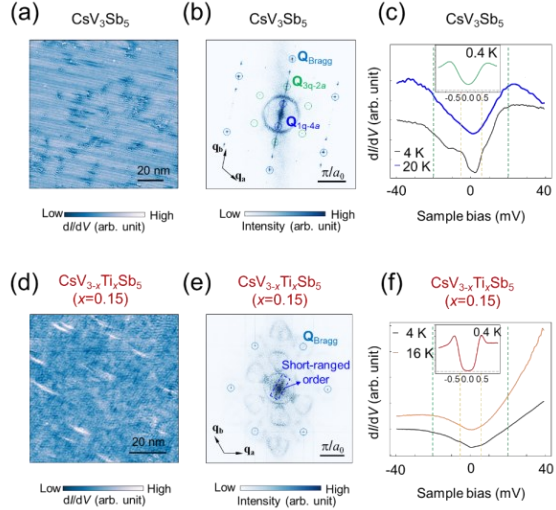

FIG 1. (a), (b) $dI/dV(r, -50\ mV)$ (a) and corresponding Fourier transform (FT) image (b) of pristine $CsV_3Sb_5$ (c) $dI/dV$ spectra of $CsV_3Sb_5$ acquired at 4 K and 20 K, respectively. Inset shows V-shape SC gap measured at 0.4 K. (d),(e) $dI/dV(r, -20\ mV)$ (d) and corresponding FT image (e) of of $CsV_{3-x}Ti_xSb_5$ ($x$=0.15). (f) $dI/dV$ spectra of $CsV_{3-x}Ti_xSb_5$ ($x$=0.15) at 4 K and 16 K. Inset shows the U-shape SC gap measured at 0.4 K.

The $CsV_3Sb_5$ single crystal displays a stacking sequence of Cs-Sb2-VSb1-Sb2-Cs layers with hexagonal symmetry (space group No.191, P 6/$mmm$) (Fig. S1(a), (b) [46]). Within the VSb1 layer, the kagome lattice of V is coordinated by Sb1 atoms at the center of the hexagons, forming a hexagonal lattice [25]. Upon Ti doping, the Ti dopants substitute V sites within the kagome layer [44]. The primary cleavage planes are the Cs and Sb2 planes (Fig. S1 [46]). On the Sb2 planes of $CsV_{3-x}Ti_xSb_5$, there are randomly distributed dark spots, corresponding to the Ti dopants in the underlying VSb1 kagome layer [44]. The substitution of V with Ti effectively tunes the cascade of exotic electronic states [24–26,28,43,44,47] in $CsV_3Sb_5$. The significant differences between the pristine and highly doped phase are directly observable through $dI/dV$ spectra and maps. The long-range 2×2 CDW and $4a_0$ charge stripes [Fig. 1(a),(b)] are significantly suppressed by Ti doping [44], changing into short-range stripes at $x$=0.15 [Fig. 1(d),(e)]. The cascade of energy scales associated with various electronic orders are examined in averaged tunneling conductance spectra. The CDW gap [18,35] around 20 meV observed in pristine $CsV_3Sb_5$ [Fig. 1(c)] persists as a gap-like feature in the Ti-doped sample [Fig. 1(f)] despite suppressed long-range CDWs. The gap of Ti-doped samples, also detected in photoemission spectroscopy and optical measurements [47], may originate from a hidden electronic order, whose onset temperature linked to a muon depolarization anomaly present in both undoped and $x$=0.15 doped compounds. At lower energies, a smaller gap emerges below ~20 K in $CsV_3Sb_5$ around 5 meV [orange dotted lines in Fig. 1(c)], identified as a pseudo-gap from the pair density wave (PDW) [25]. This pseudo-gap weakens but remains visible in the Ti-doped sample, indicating a transition to a short-range PDW. The gap maps further support the Ti-doping-induced suppression of long-range PDW (Figs. S2(c), (d) [46]). Below the superconducting (SC) transition temperature $T_c$, Ti substitution transforms the V-shaped SC gap with residual zero-energy conductance in the pristine [inset of Fig. 1(c)] and lightly-doped phase into a U-shaped SC gap without residual zero-energy conductance in highly-doped phase [inset of Fig. 1(f)]. It should be noted that the descriptions of V-shaped or U-shaped gaps based on residual zero-energy conductance at 0.4 K, without making definitive conclusions about the nodal or nodeless nature of the gap function [25,39,48].

To investigate the orbital origins of these exotic electronic states, we apply QPI imaging, a powerful technique for determining the orbital features of electronic structures in correlated materials [10,14]. We first examine theoretical results for $CsV_3Sb_5$. The calculated FS at $k_z=\pi$ mainly comprises three sheets [Fig. 2(a)] stemming from different orbitals: Sb-$p_z$ orbitals (blue), V out-of-plane $d_{xz}$ and $d_{yz}$ orbitals (green), and V in-plane $d_{xy}$ and $d_{x^2-y^2}$ orbitals (red). We perform orbital-resolved QPI simulation [Fig. 2(b)] by considering the scattering vectors within distinct orbital subbands in the constant-energy contour (CEC) [Fig. 2(a)] of the calculated bulk bands (see details in Supplemental Material [46]). Then we compare experimental QPI patterns with theoretical ones. The $dI/dV$ maps ($dI/dV(r, V)$) and corresponding Fourier transform (FT) $dI/dV(q, V)$ reveal three types of vectors: (1) Non-dispersive wave vectors of the long-range 2×2 CDW and 1×4 charge stripes [18,25,49,50] (Figs. S3 [46]). (2) Circular QPI pattern indicates an isotropic scattering vector $q_1$ [18], primarily from Sb $p_z$ orbitals [Figs. S3 [46] and Fig. 2(b)]. (3) Complex scattering wave vectors are dominated by V $d$-orbitals near the Fermi level (i.e. $dI/dV(q, -5 \text{ mV}$ [Fig. 2(c) and Fig. 2(b)]).

Figure 2(c) reveals complex patterns with unidirectional patches parallel to the 1×4 charge order ($q_2$, green dashed lines) and broken triangles with arcs perpendicular to the 1×4 charge order ($q_3$, red dashed triangles). These patches are visible between energies of ±12 mV and disappear around 35 K, reported as coherent quasiparticles in symmetry-broken electronic states [17,28]. Comparing with calculations [Fig. 2(d)], $q_2$ is attributed to scattering of out-of-plane V-$d_{xz}$ and $d_{yz}$ orbitals, with energy-dependent dispersion shown in Fig. S4 [46]; while $q_3$ stems from scattering of in-plane V-$d_{xy}$ and $d_{x^2-y^2}$ orbitals [Fig. 2(b)]. The $C_2$-symmetry in the V in-plane orbital has been reported with evidence of unidirectional electron-phonon coupling in the nematic state [51]. This indicates that electronic states from both V out-of-plane and in-plane orbitals exhibit unidirectionality, breaking the crystalline symmetry into $C_2$ symmetry.

In the superconducting states, the low-energy QPIs are dominated by Bogoliubov quasi-particles as the energy approaches the superconducting gap [Fig. 2(e)]. The FTs of $dI/dV$ maps show $q_2$ and $q_3$ QPIs [Fig. 2(e)] at the energy both beyond and within the V-shape gap [25], consistent with the finite zero-energy density of states. Additionally, The $q_3$ arcs alongside the unidirectional patches $q_2$ becomes clearer, also showing $C_2$-symmetry. Notably, the circular pattern from Sb $p_z$ orbitals $q_1$ fades and disappears below 5 mV [Fig. 2(e) and Fig. S5 [46]], coinciding with the the pseudo-gap observed [Fig. 1(c)] in CsV$_3$Sb$_5$. A radially averaged linecut in FTs of $dI/dV$ maps demonstrates the evolution of $q_1$ in the -11 mV ~ 11 mV range [Fig. 2(f)], with suppressed intensity within the -5 mV~5 mV pseudo-gap. The energy-dependent FT intensity clearly reveals a pseudo-gap feature [Fig. 2(g)]. The absence of $q_1$ within the pseudo-gap and SC gap energy ranges suggests its significant contribution to their formation.

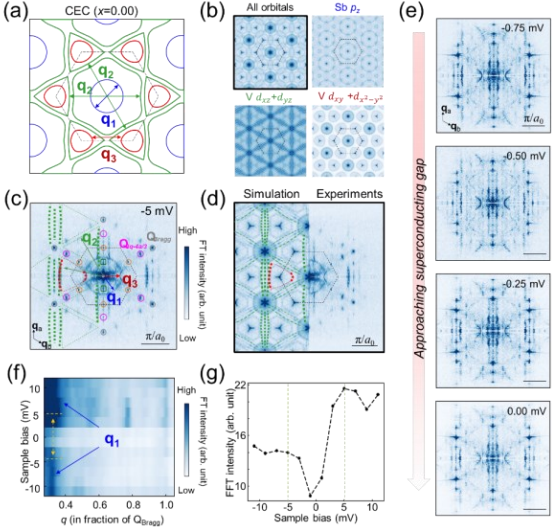

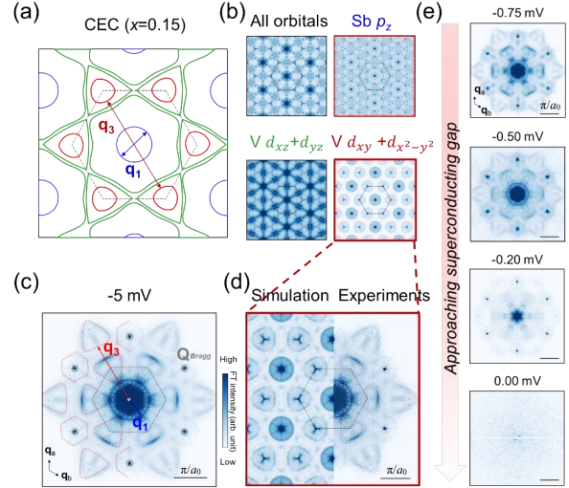

FIG 3. (a) Calculated CEC at Fermi surface of CsV$_{3-x}$Ti$_x$Sb$_5$ ($x$=0.15). (b) The Simulated QPI patterns based on all the orbitals, Sb $p_z$ orbital, V $d_{xz}$ and $d_{yz}$ orbitals and V $d_{xy}$ + $d_{x^2-y^2}$ orbitals in CEC of (a). (c) FT of $dI/dV$ map in the normal state of CsV$_{3-x}$Ti$_x$Sb$_5$ ($x$=0.15) at -5 mV. (d) The measured -5 mV QPI patterns (right) and the simulated Sb $p_z$ orbitals and V $d_{xy}$ + $d_{x^2-y^2}$ orbitals QPI pattern (left). (e) FT of $dI/dV$ map in the superconducting state of CsV$_{3-x}$Ti$_x$Sb$_5$ ($x$=0.15).

FIG. 2. (a) Calculated Fermi surface of pristine CsV$_3$Sb$_5$. The Sb $p_z$ orbital, V out-of-plane $d_{xz}$ and $d_{yz}$ and in-plane $d_{xy}$ + $d_{x^2-y^2}$ orbitals are highlighted by the blue, green and red color, respectively. (b) The simulated QPI patterns based on all the orbitals, Sb $p_z$ orbital, V $d_{xz}$ and $d_{yz}$ orbitals and V $d_{xy}$ + $d_{x^2-y^2}$ orbitals in CEC of (a). (c) FT of $dI/dV$ map in the normal state of CsV$_3$Sb$_5$ at -5 mV. (d) The measured -5 mV QPI patterns (right) and the simulated all orbitals QPI pattern (left). (e) The FT of $dI/dV$ map in the superconducting state of CsV$_3$Sb$_5$. (f) Radially averaged linecut in FTs of $dI/dV$ maps. (g) Energy-dependent FT intensity of scattering vectors $q_1$. The FT intensity is the average of 6 pixels in $q$ space at $q_1$.

Next, we perform QPI measurements in Ti-doped sample, CsV$_{3-x}$Ti$_x$Sb$_5$ ($x$=0.15). Based on the doping-induced energy shifts of Sb $p_z$ band observed in both experiment and calculation (Figs. S6, S7 [46]), we utilize a CEC [Fig. 3(a)] with an energy 40 meV lower than pristine counterpart as the FS of CsV$_{3-x}$Ti$_x$Sb$_5$ ($x$=0.15). The FS consists of Sb $p_z$ band and V 3d bands, similar to that of CsV$_3$Sb$_5$. QPI simulations, orbital by orbital, are presented in Figure 3(b). In experiments, the periodic density wave modulations and their corresponding wave vectors are completely suppressed (Figs. S8 [46]). Unlike the dominance of unidirectional coherent electronic states $q_2$ and $q_3$ in the low-energy QPI of the pristine sample, the $dI/dV(q, -5 mV)$ reveals the dominant feature $q_3$ [red dotted lines in Fig. 3(c)]. Surprisingly, only the QPI simulations from the inter-pockets and intra-pockets scattering of in-plane V-$d_{xy}$ and $d_{x^2-y^2}$ orbitals centered around the K points align with the QPI patterns of the Ti-doped sample [Fig. 3(d)]. The dispersion of $q_3$ matches DFT calculations, further confirming its origin from V-$d_{xy}$ and $d_{x^2-y^2}$ orbitals (Fig. S10 [46]). The QPI of Bogoliubov quasi-particles retains the overall patterns [Fig. 3(e)] but diminishes deep within the superconducting gap, reflecting the U-shaped superconducting gap with a depletion of the density of states. Intriguingly, the $q_1$ scattering circle vanishes around -0.2 meV, while the $q_3$ pattern remains detectable, indicating distinct orbital contributions to superconductivity in Ti-doped CsV$_3$Sb$_5$ (Fig. S11 [46]).

To track the evolution of V orbitals with Ti doping, QPI measurements on the Sb surfaces of CsV$_{3-x}$Ti$_x$Sb$_5$ with varying concentrations are performed. The low energy QPIs in lightly-substituted samples ($x$=0.03, 0.04), situated within the first superconducting dome, are primarily dominated by both $q_2$ of V out-of-plane orbital and $q_3$ of V in-plane orbitals [Fig. 4(a)], similar to the patterns observed in CsV$_3$Sb$_5$. With increasing $x$, the anisotropic stripe $q_3$ weakens, while the scattering of V in-plane $d_{xy}$ and $d_{x^2-y^2}$ orbitals $q_3$ remains [right panel of Fig. 4(a)]. The highly-substituted CsV$_{3-x}$Ti$_x$Sb$_5$ ($x$=0.15, 0.27) transitions into the second superconducting dome. Apart from $q_1$ of Sb $p_z$ orbital, their QPIs are predominantly characterized by scattering vectors $q_3$ from V in-plane orbitals [Fig. 4(b)]. Notably, the scattering vectors $q_2$ from V out-of-plane orbitals are absent in the second superconducting phase without long-range CDWs.

In general, the segmented/hexagonal FSs ($d_{xz}$ and $d_{yz}$) present stronger QPIs due to ideal nesting vectors ($q_2$) and higher density of states in the kagome lattice. Thus, the promotion of smaller triangular FSs around the K points from $d_{xy}$ and $d_{x^2-y^2}$ orbitals in QPIs by Ti doping is a striking feature. To gain a physical understanding, we performed DFT calculation of the orbital resolved charge density redistribution due to Ti substitution of V atom in CsV$_3$Sb$_5$. When a V atom is replaced by a Ti atom, the charge modulations on the nearby V atoms show a clear pattern that involves primarily the in-plane $d_{xy}$ and $d_{x^2-y^2}$ orbitals (Fig. S12 [46]). This indicates that the in-plane orbitals experience stronger scattering by the Ti dopants via the symmetry-allowed orbital hybridization than the out-of-plane $d_{xz}$ and $d_{yz}$ orbitals. This may account for the enhancement of the $q_3$ QPI due to the Fermi pockets around the K points [Figs. 4(a) and 4(b)]. However, the disappearance of $q_2$ from the out-of-plane orbitals scattering is puzzling, as Ti substitution of V suppresses the long-range CDWs and the gapped electrons of V-$d_{xz}$ and $d_{yz}$ orbitals should be released. Our DFT calculations did not reveal any suppression of out-of-plane scattering from Ti impurities. Additionally, similar impurities on the Sb surface in Ti-doped samples (Fig. S13 [46]) should also be capable of scattering out-of-plane orbitals, but was absent. This anomaly may be attributed to the hidden order that gaps the electrons [Fig. 1(e)] in the V-$d_{xz}$ and $d_{yz}$ orbitals.

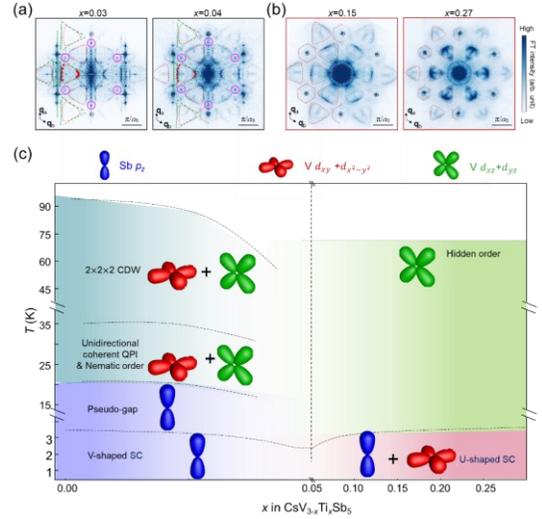

FIG 4. (a) FTs of $dI/dV(r,-5mV)$ map of CsV$_{3-x}$Ti$_x$Sb$_5$, $x$=0.03, 0.04. (b) FTs of $dI/dV(r, -5mV)$ map of CsV$_{3-x}$Ti$_x$Sb$_5$, $x$=0.15, 0.27. (c) Schematic phase diagram of CsV$_{3-x}$Ti$_x$Sb$_5$. The insets are the schematic of three orbitals labelled into each electronic phase based on QPI measurements.

The orbital-dependent QPI patterns manifest distinct physics originating from different orbitals. In CsV$_3$Sb$_5$, both the low-energy QPI of V out-of-plane and in-plane orbitals exhibit unidirectional patterns, suggesting that all the V $d$-bands undergo an orbital-dependent renormalization, thereby playing significant roles in the nematic phase. However, the V out-of-plane and in-plane orbitals contribute differently to other exotic physics of CsV$_3$Sb$_5$ system. With the suppression of CDW, the QPI patterns of V out-of-plane orbitals disappear in the Ti-doped samples, while the hidden order remains, indicating that the hidden order is primarily contributed by the V out-of-plane orbitals. Regarding the SC states, the absence of Sb $p_z$-orbital QPI inside the pseudo-gap and SC gap implies the participation of Sb $p_z$ electrons in the formation of the pseudo-gap and therefore SC states in the pristine sample. Recent works [48,52,53] suggest an anisotropic SC gap in the V orbitals and an isotropic SC gap in the Sb orbital, further emphasizing the orbital-dependent SC states. In the Ti-doped sample, however, both Sb $p_z$ orbital and V in-plane $d$ orbital QPIs appear in the Bogoliubov QPI image, gapped by the U-shape SC gap successively, indicating a multiband superconductivity.

To summarize, we construct an orbital-resolved phase diagram of Ti-doped CsV$_3$Sb$_5$ with the Ti doping contents [Fig. 4(c)]. We label the Sb $p_z$ orbital, V-$d_{xz}$ and $d_{yz}$ orbitals and V-$d_{xy}$ and $d_{x^2-y^2}$ orbitals onto different exotic electronic states based on QPI measurements. Notably, multi-orbital physics are not limited to the $A$V$_3$Sb$_5$ family but are also prominent in other kagome materials, such as orbital-selective electronic nematicity observed in $A$Ti$_3$Bi$_5$ [54,55] ($A$ = K, Rb, Cs) and ScV$_6$Sb$_6$ [56].

In conclusion, we have distinctively revealed the orbital origin of exotic electronic states in kagome superconductor CsV$_{3-x}$Ti$_x$Sb$_5$ through orbital-dependent QPI measurements across different Ti concentrations. These findings offer novel insights and avenues for exploring the crucial role of atomic orbitals in the cascade of symmetry-breaking correlated electronic states and superconductivity in kagome superconductors with broad implications in other quantum materials exhibiting multiorbital physics. Furthermore, doped CsV$_3$Sb$_5$ serves as a versatile platform for exploring exotic physics, with Cr dopants inducing magnetism [44,57], Mo dopants enhancing CDW [58], and Ta dopants introducing orbital-dependent doping effect [59].